\documentclass[conference]{IEEEtran}
\IEEEoverridecommandlockouts
\pdfoutput=1
\usepackage{cite}
\usepackage{amsmath,amssymb,amsfonts}
\usepackage{algorithmic}
\usepackage{graphicx}
\usepackage{textcomp}
\usepackage{xcolor}
\usepackage{enumitem}
\usepackage{lipsum}
\usepackage{stfloats}
\usepackage{booktabs}
\usepackage[symbol]{footmisc}
\usepackage{multirow}
\usepackage{xcolor}
\usepackage{titlesec}
\usepackage{xfp}

\def\BibTeX{{\rm B\kern-.05em{\sc i\kern-.025em b}\kern-.08em
    T\kern-.1667em\lower.7ex\hbox{E}\kern-.125emX}}

\IEEEaftertitletext{\vspace{-0.7\baselineskip}}

\usepackage[hypertexnames=false,bookmarks=false]{hyperref} 
\hypersetup{
     			colorlinks,
                linkcolor={black},
                citecolor={black},
                urlcolor={black}}
    \IEEEoverridecommandlockouts

\begin{document}
\bstctlcite{IEEEexample:BSTcontrol}

\title{A Programmable SoC for Heterogeneous\\ Neural Acquisition\vspace{-4mm}
}

\author{
\IEEEauthorblockN{
Juan Sapriza$^{1}$, Natalia Mart\'inez$^{2,3,4}$, Tommaso Terzano$^1$, David Mallas\'en$^1$, Davide Schiavone$^1$,\\
Berkay Özbek$^{2,3}$, Timothy G. Constandinou$^{2,3,4}$, and David Atienza$^1$
}

\IEEEauthorblockA{
\small
$^1$Embedded Systems Laboratory (ESL), EPFL, Switzerland, $^2$Dept. of Electrical and Electronic Engineering, Imperial College London, UK, \\
$^3$UK Dementia Research Institute, Care Research \& Technology Centre, UK, $^4$Mint Neurotechnologies Ltd, UK \\
Email: \href{mailto:juan.sapriza@epfl.ch}{juan.sapriza@epfl.ch},
\href{mailto:n.martinez23@imperial.ac.uk}{n.martinez23@imperial.ac.uk} \vspace{-3mm}
}

\thanks{
\hspace*{-2\parindent}
\begin{minipage}[t]{\linewidth}
This work was supported in part by the Swiss State Secretariat for
Education, Research, and Innovation (SERI) through the SwissChips
research project, and in part by the Imperial College London President's
PhD Scholarship.
\end{minipage}
}
}

\newcommand{\PwrPathMin}{{\color{red}1234}}
\newcommand{\PwrPathMax}{{\color{red}1234}}
\newcommand{\RatePathMin}{{\color{red}1234}}
\newcommand{\RatePathMax}{{\color{red}1234}}
\newcommand{\BackendMaxInputRate}{{\color{red}1234}}
\newcommand{\BackendMixedPower}{{\color{red}1234}}

\newcommand{\LNAArea}{{\color{black}44,000}}
\newcommand{\LNAPowerVal}{2.3}
\newcommand{\LNAPower}{{\color{black}\LNAPowerVal}}
\newcommand{\LNABandwidth}{{\color{black}5\,m--5\,k Hz}}

\newcommand{\BPEArea}{{\color{black}287,000}}
\newcommand{\BPEPowerVal}{0.5}
\newcommand{\BPEPower}{{\color{black}\BPEPowerVal}}
\newcommand{\BPEBandwidth}{{\color{black}10--126\,Hz}}

\newcommand{\DSMArea}{{\color{black}150,000}}
\newcommand{\DSMPowerLowVal}{4.7}
\newcommand{\DSMPowerHighVal}{12.6}
\newcommand{\DSMPower}{{\color{black}\DSMPowerLowVal/\DSMPowerHighVal}}
\newcommand{\DSMMaxRate}{{\color{black}50/1,000\,kbps}}

\newcommand{\VCOArea}{{\color{black}28,800}}
\newcommand{\VCOPowerLowVal}{1.0}
\newcommand{\VCOPowerHighVal}{5.0}
\newcommand{\VCOPower}{{\color{black}\VCOPowerLowVal--\VCOPowerHighVal}}
\newcommand{\VCOMaxRate}{{\color{black}100--1,200\,kbps}}

\newcommand{\LCArea}{{\color{black}38,750}}
\newcommand{\LCPowerLowVal}{1.5}
\newcommand{\LCPowerHighVal}{2.5}
\newcommand{\LCPower}{{\color{black}\LCPowerLowVal--\LCPowerHighVal}}
\newcommand{\LCMaxRate}{{\color{black}4.2--2,500\,kbps}} %

\newcommand{\CICArea}{{\color{black}2,395 /stage}}
\newcommand{\CICPowerPerStageVal}{0.8}
\newcommand{\CICPower}{{\color{black}\CICPowerPerStageVal/stage}}
\newcommand{\CICMaxRate}{{\color{black}80--1,000\,kbps}}

\newcommand{\SESArea}{{\color{black}1,205 /stage}}
\newcommand{\SESPowerPerStageVal}{0.7}
\newcommand{\SESPower}{{\color{black}\SESPowerPerStageVal/stage}}
\newcommand{\SESMaxRate}{{\color{black}80--1,000\,kbps}}

\newcommand{\VCOCounterArea}{{\color{black}6,220}}
\newcommand{\VCOCounterPowerLowVal}{1.0}
\newcommand{\VCOCounterPowerHighVal}{4.1}
\newcommand{\VCOCounterPower}{{\color{black}\VCOCounterPowerLowVal--\VCOCounterPowerHighVal}}
\newcommand{\VCOCounterMaxRate}{{\color{black}0--16\,kbps}}

\newcommand{\DLCArea}{{\color{black}6,501}}
\newcommand{\DLCPowerVal}{1.31}
\newcommand{\DLCPower}{{\color{black}\DLCPowerVal}}
\newcommand{\DLCMaxRate}{{\color{black}39\,bps--500\,kbps}} %

\newcommand{\BackendClock}{1\,MHz}
\newcommand{\CPUArea}{{\color{black}33,336}}
\newcommand{\CPUPowerLowVal}{1.2}
\newcommand{\CPUPowerHighVal}{6.9}
\newcommand{\CPUPower}{{\color{black}\CPUPowerLowVal--\CPUPowerHighVal}}
\newcommand{\CPUclock}{\color{black}{$<$50\,MHz}}

\newcommand{\CtrlArea}{{\color{black}$<$568,202}}
\newcommand{\CtrlPowerMaxVal}{20}
\newcommand{\CtrlPower}{{\color{black}$<$\CtrlPowerMaxVal}}
\newcommand{\CtrlClock}{\color{black}{$>$1\,MHz}}

\newcommand{\FunnelArea}{{\color{black}2,282}}
\newcommand{\FunnelPowerVal}{0.18}
\newcommand{\FunnelPowerMax}{{\color{black}\FunnelPowerVal}}
\newcommand{\FunnelMaxRate}{{\color{black}16\,Mbps/MHz}}

\newcommand{\DMAArea}{{\color{black}25,605}}
\newcommand{\DMAPowerLowVal}{0.8}
\newcommand{\DMAPowerHighVal}{2.1}
\newcommand{\DMAPowerMax}{{\color{black}\DMAPowerLowVal--\DMAPowerHighVal}}
\newcommand{\DMAMaxRate}{{\color{black}64\,Mbps/MHz}}

\newcommand{\BusArea}{{\color{black}5,826}}
\newcommand{\BusPowerVal}{0.3}
\newcommand{\BusPowerMax}{{\color{black}\BusPowerVal}}
\newcommand{\BusMaxRate}{{\color{black}64\,Mbps/MHz}}

\newcommand{\MemoryArea}{{\color{black}101,301}}
\newcommand{\MemoryPowerLowVal}{6.1}
\newcommand{\MemoryPowerHighVal}{47.4}
\newcommand{\MemoryPowerMax}{{\color{black}\MemoryPowerLowVal--\MemoryPowerHighVal}}
\newcommand{\MemoryMaxRate}{{\color{black}64\,Mbps/MHz}}

\newcommand{\ParallelArea}{{\color{black}88}}
\newcommand{\ParallelPowerVal}{11.6} %
\newcommand{\ParallelPower}{{\color{black}\ParallelPowerVal}}
\newcommand{\ParallelMaxRate}{{\color{black}8\,Mbps/MHz}}

\newcommand{\SPIArea}{{\color{black}17,224}}
\newcommand{\SPIPower}{{\color{black}12.5}}
\newcommand{\SPIMaxRate}{{\color{black}0.7\,Mbps/MHz}}

\newcommand{\WirelessULArea}{{\color{black}4,000}}
\newcommand{\WirelessULPowerVal}{0.8}
\newcommand{\WirelessULPower}{{\color{black}\WirelessULPowerVal}}
\newcommand{\WirelessULMaxRate}{{\color{black}0.5\,Mbps}}

\newcommand{\WirelessDLArea}{{\color{black}102,600}}
\newcommand{\WirelessDLPowerVal}{12.6}
\newcommand{\WirelessDLPower}{{\color{black}\WirelessDLPowerVal}}
\newcommand{\WirelessDLMaxRate}{{\color{black}0.2\,Mbps}}

\newcommand{\ClockRecoveryArea}{{\color{black}1,380}}
\newcommand{\ClockRecoveryPowerVal}{3.1}
\newcommand{\ClockRecoveryPower}{{\color{black}\ClockRecoveryPowerVal}}
\newcommand{\ClockRecoveryFreq}{{\color{black}1\,MHz}}

\newcommand{\CurrentRefArea}{{\color{black}76,000}}
\newcommand{\CurrentRefPowerVal}{1.1}
\newcommand{\CurrentRefPower}{{\color{black}\CurrentRefPowerVal}}
\newcommand{\CurrentRefRange}{{\color{black}$20\times 1$\,\textmu A}}

\newcommand{\PowerManagementArea}{{\color{black}319,990}}
\newcommand{\PowerManagementPowerLowVal}{34.8}
\newcommand{\PowerManagementPowerHighVal}{72.5}
\newcommand{\PowerManagementPower}{{\color{black}\PowerManagementPowerLowVal--\PowerManagementPowerHighVal}}
\newcommand{\PowerManagementMaxLoad}{{\color{black}0.4/0.5/0.6/1.0\,V}}

\newcommand{\SoCArea}{{\color{black}5.2\,mm$^2$}}          %
\newcommand{\SoCPower}{{\color{red}1234}}
\newcommand{\SoCPerformance}{{\color{red}1234}}          %

\newcommand{\CoreArea}{{\color{black}2.85\,mm$^2$}}         %

\newcommand{\UseCaseABasePowerVal}{\fpeval{
    3*\LNAPowerVal
    + 2*\DSMPowerHighVal
    + 2*3*\CICPowerPerStageVal
    + 3*\LCPowerHighVal
    + 3*\DLCPowerVal
    + \FunnelPowerVal
    + \DMAPowerHighVal
    + \BusPowerVal
    + \MemoryPowerLowVal
    + \ParallelPowerVal
    + \CPUPowerLowVal
    + \ClockRecoveryPowerVal
    + \CurrentRefPowerVal
}}
\newcommand{\UseCaseATotalPowerLowVal}{\fpeval{\UseCaseABasePowerVal+\PowerManagementPowerLowVal}}
\newcommand{\UseCaseATotalPowerHighVal}{\fpeval{\UseCaseABasePowerVal+\PowerManagementPowerHighVal}}
\newcommand{\UseCaseAPowerPerChannel}{\fpeval{round(\UseCaseATotalPowerHighVal/5,0)}}

\newcommand{\UseCaseBBasePowerVal}{\fpeval{
    3*\LNAPowerVal
    + \BPEPowerVal
    + 2*\DSMPowerLowVal
    + 2*3*\SESPowerPerStageVal
    + 3*\VCOPowerHighVal
    + 3*\VCOCounterPowerHighVal
    + \FunnelPowerVal
    + \DMAPowerHighVal
    + \BusPowerVal
    + \MemoryPowerLowVal
    + \WirelessULPowerVal
    + \CPUPowerLowVal
    + \ClockRecoveryPowerVal
    + \CurrentRefPowerVal
}}
\newcommand{\UseCaseBTotalPowerLowVal}{\fpeval{\UseCaseBBasePowerVal+\PowerManagementPowerLowVal}}
\newcommand{\UseCaseBTotalPowerHighVal}{\fpeval{\UseCaseBBasePowerVal+\PowerManagementPowerHighVal}}
\newcommand{\UseCaseBPowerPerChannel}{\fpeval{round(\UseCaseBTotalPowerHighVal/5,0)}}

\newcommand{\UseCaseCBasePowerVal}{\fpeval{
    2*\LNAPowerVal
    + \BPEPowerVal
    + \LCPowerLowVal
    + \LCPowerHighVal
    + 2*\DLCPowerVal
    + \FunnelPowerVal
    + \DMAPowerHighVal
    + \BusPowerVal
    + \MemoryPowerHighVal
    + \WirelessULPowerVal
    + \CPUPowerHighVal
    + \ClockRecoveryPowerVal
    + \CurrentRefPowerVal
}}
\newcommand{\UseCaseCTotalPowerLowVal}{\fpeval{\UseCaseCBasePowerVal+\PowerManagementPowerLowVal}}
\newcommand{\UseCaseCTotalPowerHighVal}{\fpeval{\UseCaseCBasePowerVal+\PowerManagementPowerHighVal}}
\newcommand{\UseCaseCPowerPerChannel}{\fpeval{\fpeval{round(\UseCaseCTotalPowerHighVal/2,0)}}}

\maketitle

\begin{abstract}

Neural interfaces are used across a wide range of applications, each with different requirements for signal bandwidth, information content, processing, and communication. However, implantable electronics are severely constrained in power and area, making a one-size-fits-all acquisition approach inefficient when only reduced-bandwidth signals, features, or sparse events are needed. This work explores heterogeneous acquisition with a shared programmable infrastructure. HEEPokranios, a \mbox{65-nm} system-on-chip (SoC), integrates delta-sigma, voltage-controlled oscillator, and level-crossing acquisition paths with analog band-power extraction and a common 32-bit \mbox{RISC-V} back-end. Dedicated processing, event handling, DMA-based data movement, buffering, and wired or wireless communication support different neural-data representations. The shared acquisition funnel occupies 2,282\,\textmu m$^2$ (0.08\% of the active core) and consumes 0.18\,\textmu W at 1\,MHz. Representative end-to-end paths are simulated using artifact-laden non-human primate microwire neural recordings. Example configurations include a 5-channel wired mode producing 641\,kbps at \UseCaseAPowerPerChannel\,\textmu W/ch, a 5-channel wireless mode producing 33.8\,kbps at \UseCaseBPowerPerChannel\,\textmu W/ch, and a 2-channel local-processing mode operating at \UseCaseCPowerPerChannel\,\textmu W/ch without continuous off-chip transmission. Simulation results illustrate the value of integrating heterogeneous acquisition strategies within a shared programmable system to address diverse neural-interface requirements.

\end{abstract}

\medskip
{}
\begin{IEEEkeywords}
Implantable neural interfaces, neural recording, heterogeneous acquisition, event-based acquisition, RISC-V.
\end{IEEEkeywords}

\section{Introduction}\label{sec:introduction}

\begin{figure}[t!]
\centering
\includegraphics[width=0.9\linewidth]{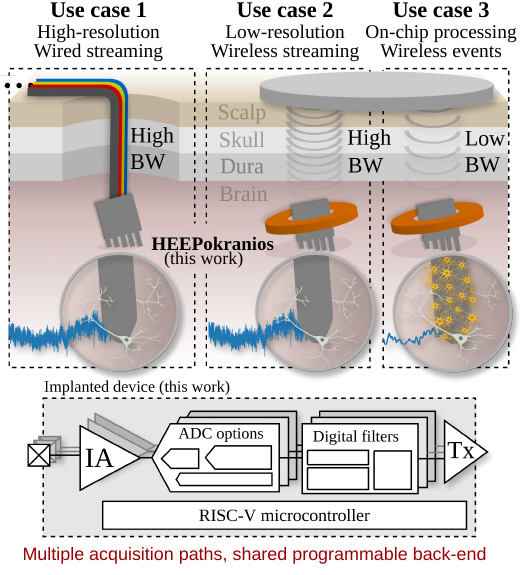}
\caption{Conceptual organization of HEEPokranios. Representative use cases show different system requirements. Complementary neural acquisition front-ends target these requirements, while a shared programmable RISC-V subsystem provides configuration, processing, data movement, and communication.}
\label{fig:concept}
\end{figure}

\IEEEPARstart{F}{rom} high-density electrophysiology to brain--computer interfaces (BCIs) and closed-loop neuromodulation, neural applications impose different requirements on electronic systems~\cite{Luan2020NeuralInterfaces}. Basic-neuroscience platforms prioritize broadband, high-resolution recordings~\cite{SteinmetzNeuropixels2Four2021}, while fully implantable BCIs should balance information preservation against power and wireless-throughput constraints~\cite{SimeralBrainGateWireless2021,MuskNeuralink2019SystemB,JungBISC2025}. Therapeutic systems may instead rely on specific biomarkers, such as beta-band power for adaptive deep brain stimulation (DBS)~\cite{Michmizos2015Beta}. These contrasting requirements are illustrated in \autoref{fig:concept}.

Such diversity is challenging in implantable electronics, where power and silicon area are severely constrained. A one-size-fits-all approach can lead to overengineered systems, allocating resources to capabilities that are not required by every channel or application. Different acquisition approaches have been developed to address these requirements, including conventional amplifier-based recording~\cite{HarrisonCharles2003,Wattanapanitch2007},
delta-sigma conversion~\cite{ChandrakumarMarkovic2018},
voltage-controlled oscillator (VCO)-based
acquisition~\cite{Muller2015uECoG}, and event-driven
approaches~\cite{Aziz2009Delta,CorradiIndiveri2015,Kim2019DeltaNeural}.
Spike detection and feature extraction can further
reduce processing and transmission
requirements~\cite{Chae2009,Verma2010SeizureSoC,Shoaran2015,AnNeuRAD2022,Maslik2018ContinuousTime}.

Despite these differences, acquisition architectures require common functions for configuration, processing, buffering, data movement, and communication. Sharing this infrastructure offers an alternative to replicating independent digital readout subsystems. However, uniformly sampled analog-to-digital converters (ADCs) generate continuous data streams, level-crossing (LC) converters generate asynchronous events, and VCO-based circuits encode information in frequency or timing. A shared back-end must therefore efficiently accommodate different timing models, data formats, rates, and processing requirements while preserving the benefits of each acquisition approach.

To explore this opportunity, we present HEEPokranios, a \mbox{65-nm} SoC integrating delta-sigma modulation (DSM)-, VCO- and LC-based acquisition, analog band-power extraction, and a shared programmable \mbox{RISC-V} back-end. Dedicated processing and event handling accommodate heterogeneous representations within common storage, data-movement, and communication resources. Representative configurations are evaluated using non-human primate recordings and post-layout power estimates to investigate application-dependent trade-offs. 
Overall, this work illustrates how heterogeneous acquisition with shared digital infrastructure can balance application-specific information requirements against constrained system resources, without relying on one-size-fits-all architectures.

\section{System Architecture and Implementation}\label{sec:architecture}

\begin{figure*}[t!]
\centering
\includegraphics[width=\linewidth]{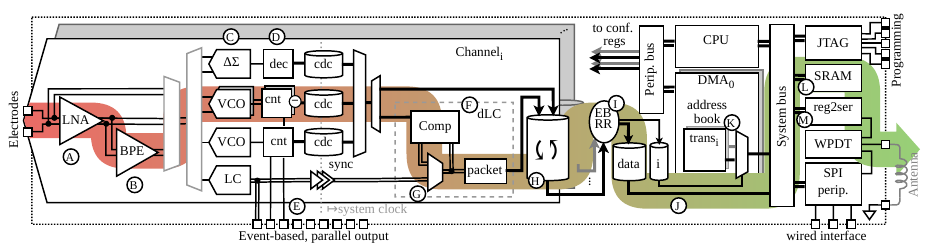}
\caption{Top-level architecture of HEEPokranios. Heterogeneous acquisition channels connect to a common X-HEEP-based digital back-end containing dedicated post-processing, event funneling, DMA, memory, and communication resources. An example data flow is highlighted. Analog multiplexing is done off-chip.}
\label{fig:architecture}
\end{figure*}

The architecture in \autoref{fig:architecture} combines complementary acquisition front-ends with a programmable digital back-end, retaining representation-specific processing while sharing data-movement, storage, and communication resources.

\subsection{Heterogeneous Acquisition Paths}

Neural signals may first be conditioned by the low-noise amplifier (LNA)~\textcircled{\scalebox{0.7}A}. For waveform acquisition, the DSM supports extracellular action potential (EAP) and local field potential (LFP) bandwidths, followed by configurable cascaded integrator-comb (CIC) or simple-exponential-smoothing (SES) filtering~\cite{martinezReconfigurableHighDynamicRange2025}. The CIC provides decimation for waveform transmission, while the SES offers lightweight filtering for local processing. 

Alternatively, VCO-based conversion maps amplitude to frequency, with timing information accumulated for waveform reconstruction or retained as counter-overflow events. LC conversion, on the other hand, generates asynchronous events from signal crossings, with finer levels supporting waveform reconstruction and coarser levels producing sparser representations. The analog band-power extractor (BPE)~\textcircled{\scalebox{0.7}B}~\cite{Ge2026BioCAS} further reduces information before VCO or LC conversion, enabling acquisition of spectral-power features. Together, these approaches form the heterogeneous acquisition stage~\textcircled{\scalebox{0.7}C}.

\subsection{Data Funneling}

Heterogeneous outputs must converge onto the shared back-end despite differing formats, timing, and data rates. After channel-specific processing~\textcircled{\scalebox{0.7}D}, the data is temporarily buffered. VCO counter-overflow and LC crossing/direction events can alternatively be exposed through dedicated digital outputs~\textcircled{\scalebox{0.7}E}. When asynchronous information must instead be stored or processed internally, the digital level-crossing (dLC) encoder~\textcircled{\scalebox{0.7}F} provides a common event representation containing the signal transition and a timestamp quantized from the system clock. LC crossing/direction events reuse this packetization infrastructure after synchronization to the system clock~\textcircled{\scalebox{0.7}G}.

Synchronized outputs are stored in a channel-wide FIFO~\textcircled{\scalebox{0.7}H}. An event-based round-robin arbiter~\textcircled{\scalebox{0.7}I} services channels containing valid data and forwards the selected word together with its channel identifier to the acquisition DMA~\textcircled{\scalebox{0.7}J}. Each channel independently defines its destination and addressing mode~\textcircled{\scalebox{0.7}K}, allowing data to be moved to SRAM~\textcircled{\scalebox{0.7}L} or directly to another peripheral (e.g., for transmission~\textcircled{\scalebox{0.7}M}). This shared infrastructure supports both direct streaming and local processing or buffering before transmission, without requiring separate data-movement hardware for each acquisition representation.

\subsection{Programmable Digital Back-End}

The shared digital subsystem is based on X-HEEP~\cite{machetti2025xheep} and integrates a 32-bit RISC-V CPU (cv32e20~\cite{davideschiavoneSlowSteadyWins2017}), 32\,kB of SRAM, two configurable DMA units, standard peripherals, and dedicated acquisition hardware. Acquisition and processing blocks are memory mapped and run-time configurable. Dedicated hardware performs decimation, VCO counting, event compression, packetization, and buffering. DMA transfers data independently of CPU execution, leaving the CPU available for application-level control and local processing.

\subsection{Wireless Power and Communication}

A dedicated digital interface supports the custom bidirectional wireless protocol and remote programming of the on-chip memory. Wired access is alternatively provided through JTAG and SPI. Uplink communication uses Manchester-encoded load modulation, with a simulated bit error rate (BER) below $1.48 \times 10^{-4}$ at a 2.2\% modulation index and a 1\,Mbps Amplitude-Shift-Keying (ASK) line rate, corresponding to a 0.5\,Mbps effective data rate. Downlink communication uses amplitude modulation of a 433\,MHz carrier, with Manchester encoding for data recovery. The same RF carrier is used for wireless power transfer~\cite{berkayCICC2026}, while dedicated cold-start circuitry and an on-chip clock generator support autonomous system initialization and operation.

\subsection{Physical Implementation and Hardware Resources}

\begin{figure}[t!]
\centering
\includegraphics[width=0.85\linewidth]{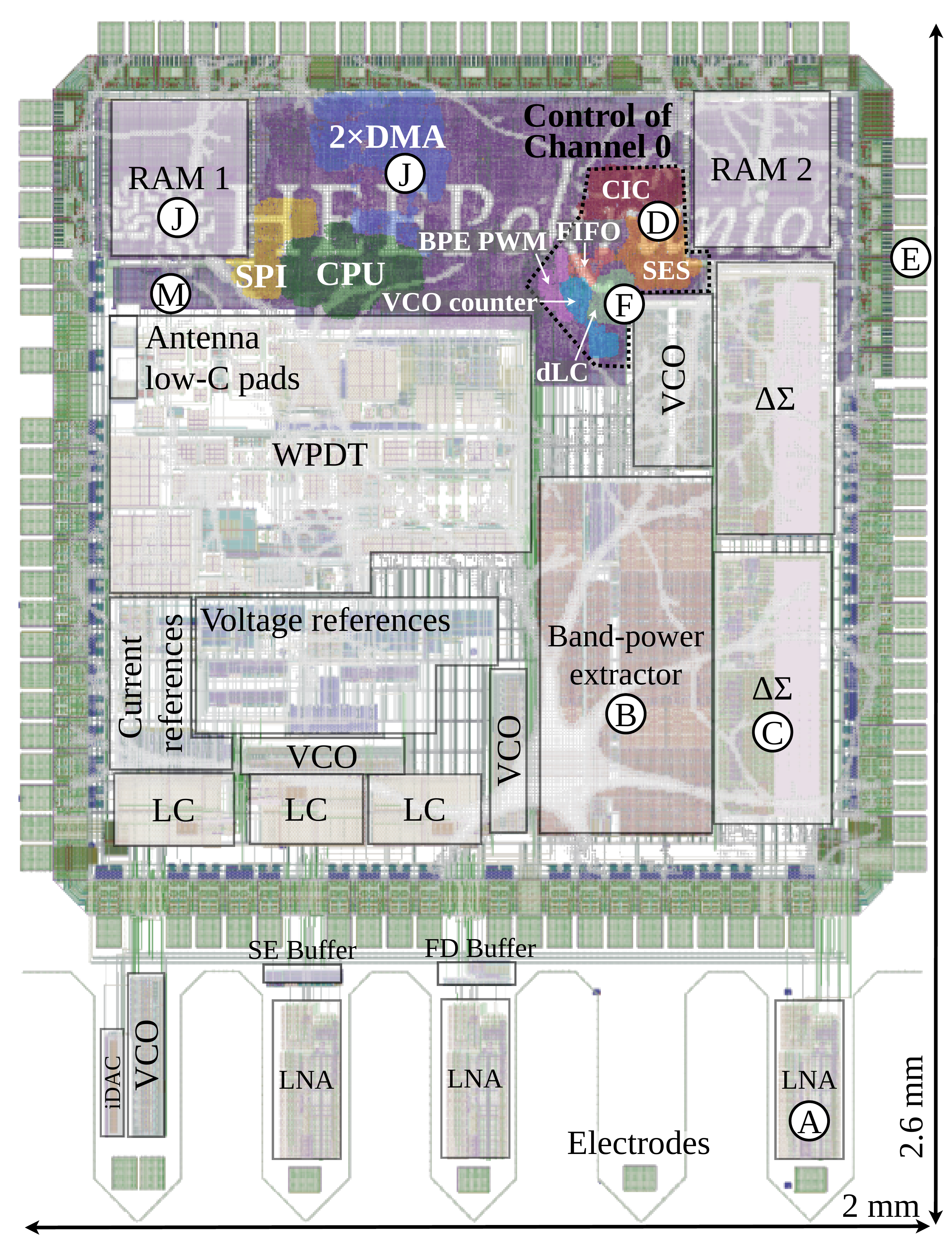}
\caption{Top-level physical implementation of HEEPokranios, including the programmable digital subsystem, SRAM banks, heterogeneous acquisition circuits, analog support blocks, wireless power/data-transfer interface, and electrode connections.}
\label{fig:layout}
\end{figure}

HEEPokranios is implemented in a 65-nm CMOS technology under a 1\,V supply and is currently under fabrication. The layout is shown in \autoref{fig:layout}, with hardware resources summarized in \autoref{tab:shared}. The SoC occupies \SoCArea, including an active core of \CoreArea, and integrates heterogeneous acquisition circuits, an impedance-monitoring 8-bit iDAC, and programmable voltage/current references.

The front-ends are implemented as dedicated test channels rather than dynamically selectable architectures per electrode. After digitization, processing, routing, storage, and communication are run-time configurable. The shared acquisition funnel occupies 2,282\,\textmu m$^2$ (0.08\% of the active core) and consumes 0.18\,\textmu W at 1\,MHz, quantifying the small convergence overhead for heterogeneous data representations.

\begin{table}[t!]
\centering
\caption{Hardware resources: area, power, and performance.}
\label{tab:shared}
\renewcommand{\arraystretch}{1}
\setlength{\tabcolsep}{3pt}
\begin{tabular}{lrrr}
\toprule
\textbf{Resource} & \textbf{Area (\textmu m$^2$)} & \textbf{Power$^*$ (\textmu W)} & \textbf{Performance} \\
\midrule

\multicolumn{3}{l}{\textit{Signal conditioning}} & \textit{Bandwidth} \\
LNA & \LNAArea & \LNAPower & \LNABandwidth \\
BPE & \BPEArea & \BPEPower & \BPEBandwidth \\

\midrule
\multicolumn{3}{l}{\textit{Digitization}} & \textit{Output data rate} \\
DSM & \DSMArea & \DSMPower & \DSMMaxRate \\
VCO ADC & \VCOArea & \VCOPower & \VCOMaxRate \\
LC & \LCArea & \LCPower & \LCMaxRate \\

\midrule
\multicolumn{3}{l}{\textit{Digital processing}} & \textit{Output data rate} \\
CIC filter ($\leq$5 stages) & \CICArea & \CICPower & \CICMaxRate \\
SES filter ($\leq$10 stages) & \SESArea & \SESPower & \SESMaxRate \\
VCO counter ($\leq$32 bits) & \VCOCounterArea & \VCOCounterPower & \VCOCounterMaxRate \\
dLC compression & \DLCArea & \DLCPower & \DLCMaxRate \\

\midrule
\multicolumn{3}{l}{\textit{Common digital back-end}} & \textit{Clock / throughput} \\
RISC-V CPU & \CPUArea & \CPUPower & \CPUclock \\
AFE controllers (all ch.) & \CtrlArea & \CtrlPower & \CtrlClock \\
Acquisition funnel & \FunnelArea & \FunnelPowerMax & \FunnelMaxRate \\
DMA & \DMAArea & \DMAPowerMax & \DMAMaxRate \\
OBI interconnect & \BusArea & \BusPowerMax & \BusMaxRate \\
SRAM (2$\times$16\,kB) & \MemoryArea & \MemoryPowerMax & \MemoryMaxRate \\

\midrule
\multicolumn{3}{l}{\textit{Transmission}} & \textit{Max. data rate} \\
8-bit parallel bus & \ParallelArea & \ParallelPower$^{\dagger\ddagger}$ & \ParallelMaxRate \\
Serial interface & \SPIArea & \SPIPower$^{\dagger\ddagger}$ & \SPIMaxRate \\
Wireless uplink & \WirelessULArea & \WirelessULPower$^{\dagger}$ & \WirelessULMaxRate \\
Wireless downlink & \WirelessDLArea & \WirelessDLPower$^{\dagger}$ & \WirelessDLMaxRate \\

\midrule
\multicolumn{3}{l}{\textit{Power, clock, and references}} & \textit{Output / range} \\
Clock recovery & \ClockRecoveryArea & \ClockRecoveryPower & \ClockRecoveryFreq \\
Current reference & \CurrentRefArea & \CurrentRefPower & \CurrentRefRange \\
Power management & \PowerManagementArea & \PowerManagementPower & \PowerManagementMaxLoad \\

\midrule
\textbf{Active core} & \CoreArea & -- & -- \\
\bottomrule
\end{tabular}

\vspace{1mm}
\parbox{\columnwidth}{\scriptsize
\textsuperscript{*}Post-layout results. \,\,
$^{\dagger}$Power at maximum data rate, at 1\,MHz. 
$^{\ddagger}$External clock and IO power not considered.
}
\end{table}

\section{Representative Configurations and Evaluation}
\label{sec:evaluation}

\begin{figure*}[t!]
\centering
\includegraphics[width=\linewidth]{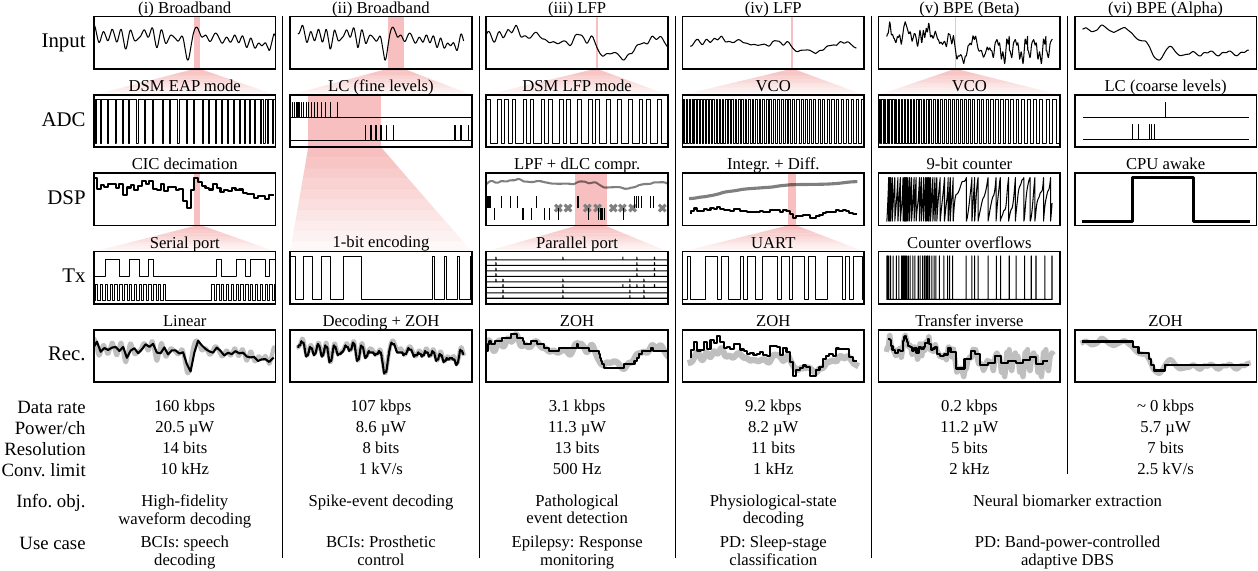}
\caption{Representative end-to-end acquisition configurations, from broadband waveforms to analog feature extraction and sparse event encoding. Artifact-laden non-human primate microwire recordings are processed through behavioral models to generate the illustrative waveforms, while power is estimated from post-layout simulations. The effective number of bits (ENOB) of each acquisition chain is reported as the resolution, together with its conversion limit (i.e., the maximum supported signal bandwidth for DSM- and VCO-based acquisition, or the maximum input slew rate for LC conversion).}
\label{fig:paths}
\end{figure*}

\autoref{fig:paths} evaluates six acquisition configurations targeting different information requirements: EAP waveforms and events (i--ii) for speech decoding~\cite{willettHighperformanceSpeechNeuroprosthesis2023} and prosthetic control~\cite{hochbergNeuronalEnsembleControl2006}; LFP acquisition (iii--iv) for seizure monitoring~\cite{ilyasForecastingSeizureClusters2022} and physiological-state decoding~\cite{yinGeneralizedSleepDecoding2024}; and band-power extraction (v--vi) for biomarker-driven DBS~\cite{stanslaskiSensingDataMethodology2024}. Non-human primate recordings are propagated through behavioral models, with power obtained from post-layout simulations.

The configurations expose distinct limiting factors. For broadband acquisition, LC (ii) reduces power by 58\% and data rate by 33\% relative to DSM (i), at the expense of lower resolution and input slew-rate constraints. For LFP acquisition, VCO (iv) consumes 27\% less power than DSM (iii), but generates approximately three times more data. When only neural features are required, BPE-based paths (v--vi) reduce information before conversion, with BPE--LC offering lower power and sparse, signal-dependent events compared with BPE--VCO.

These results show that the preferred acquisition path will depend on the dominant constraint: power, bandwidth, resolution, or information requirements. Heterogeneity does not lower per-AFE power; instead it exposes complementary operating points that let the system match acquisition cost to information need.

The system configurations in \autoref{tab:sota}, corresponding to use cases 1--3 in \autoref{fig:concept}, illustrate how these trade-offs translate to system-level operation. Use case~1 combines two high-resolution DSM and three lower-resolution LC broadband paths. This configuration can support long-term neuroscience studies requiring full-bandwidth EAP recordings for external processing, while accommodating channels with different signal quality (e.g., due to the evolution of tissue response and scar formation~\cite{woeppel2021explant,erbsloh2024modern}). Its aggregate data rate exceeds the 0.5\,Mbps wireless uplink capacity, motivating wired communication.

Use case~2 instead targets applications requiring primarily LFP-bandwidth recordings, complemented by band-power spectral features. Combining two LFP DSM and three VCO paths, it generates 33.8\,kbps at \UseCaseBPowerPerChannel\,\textmu W/ch. Despite comparable power per channel, this represents a $19\times$ lower data rate than use case~1, reflecting different information requirements. It occupies only 6.8\% of the wireless uplink capacity, illustrating how acquisition flexibility enables untethered operation under bandwidth constraints. In both cases, the shared digital funnel, supporting 16\,Mbps/MHz, can comfortably accommodate the heterogeneous data streams. Beyond communication bandwidth, silicon area also constrains scalability. A channel with LNA--BPE--DSM occupies approximately 0.48\,mm$^2$ (17\% of the active core), with BPE and DSM accounting for approximately 90\% of this area, compared with only 0.112\,mm$^2$ for an LNA--VCO--LC path. Although the prototype uses dedicated test channels, future implementations could selectively distribute acquisition resources rather than uniformly replicating all architectures.

Use case~3 explores an untethered platform for neuroscience experiments requiring local processing rather than continuous data streaming. Two LC-based channels operate at \UseCaseCPowerPerChannel\,\textmu W/ch, allowing neural information to be processed on-chip. The higher power per channel primarily reflects the shared platform overhead being distributed across only two active channels, rather than an intrinsic penalty of LC acquisition.

\begin{table}[b!]
\centering
\caption{Representative neural-interface operating regimes.}
\label{tab:sota}
\scriptsize
\setlength{\tabcolsep}{2.0pt}
\renewcommand{\arraystretch}{1.12}
\begin{tabular*}{\columnwidth}{@{\extracolsep{\fill}}lccc@{}}
\toprule
\textbf{Work} &
\textbf{Signal} &
\textbf{ADC channels} &
\shortstack{\textbf{Power/ch}$^*$ \textbf{($\mu$W)}}\\
\midrule

\multicolumn{4}{@{}l}{\textit{Broadband streaming}} \\
Reich et al. \cite{reichHighVoltageCompliance32Channel2021}
& EAP+LFP & 32$\times$16-b DSM & 193$^\dagger$ \\
Yeon et al. \cite{yeonDistributedFreefloatingWireless2016}
& EAP & 1$\times$10-b VCO & $93$$^\dagger$ \\
This work (Use case 1)
& EAP & 2$\times$14-b DSM + 3$\times$8-b LC
& \UseCaseAPowerPerChannel$^\dagger$ \\

\midrule
\multicolumn{4}{@{}l}{\textit{Reduced-data streaming}} \\
Uran et al. \cite{uran16ChannelNeuralRecording2022}
& EAP+LFP & 16$\times$10-b SAR & 25 \\
Lim et al. \cite{limLightTolerantWirelessNeural2022}
& SBP & 1$\times$2.8-b Flash & 0.57 \\
This work (Use case 2)
& BPE+LFP & 2$\times$13-b DSM + 3$\times$11-b VCO
& \UseCaseBPowerPerChannel$^\dagger$ \\

\midrule
\multicolumn{4}{@{}l}{\textit{On-chip processing}} \\
Cheng et al. \cite{chengFullyIntegrated16Channel2018}
& ECoG & 16$\times$10-b SAR & 195 \\
Guo et al. \cite{guo68channelNeuralSignal2024}
& EAP/LFP & 68$\times$9-b DSM & 0.41$^\ddagger$\\
This work (Use case 3)
& BPE+EAP & 1$\times$7-b LC + 1$\times$8-b LC
& \UseCaseCPowerPerChannel$^\dagger$ \\

\bottomrule
\end{tabular*}

\vspace{0.5mm}
\parbox{\columnwidth}{\scriptsize
$^*$Power/ch: complete system power divided by the number of concurrent channels. \\ $^\dagger$ Simulation results. $^\ddagger$ Excludes compression, telemetry, and power management.
}
\end{table}

The state-of-the-art comparison in \autoref{tab:sota} highlights different strategies for addressing neural-interface constraints. Broadband systems~\cite{reichHighVoltageCompliance32Channel2021,yeonDistributedFreefloatingWireless2016} prioritize EAP waveforms, whereas reduced-data approaches~\cite{uran16ChannelNeuralRecording2022,limLightTolerantWirelessNeural2022} use lower-resolution or compressed representations. Other architectures~\cite{chengFullyIntegrated16Channel2018,guo68channelNeuralSignal2024} incorporate on-chip processing to limit raw-data transmission. HEEPokranios accommodates these operating regimes within a shared programmable back-end, achieving competitive power consumption and output data rates. Although the prototype integrates only a few channels, increasing the channel count could amortize the shared system power overhead. Moreover, as shown in Table~I, lower-resolution acquisition paths occupy less silicon area, allowing channel count to scale without proportionally increasing area when high-resolution recordings are not required.

\section{Conclusion}\label{sec:conclusion}

This work presented HEEPokranios, a 65-nm SoC integrating heterogeneous neural acquisition with a shared programmable RISC-V back-end. Representative configurations highlight the competing trade-offs between acquisition power, information preservation, processing, and communication bandwidth, with no single architecture being optimal across all use cases. Different constraints therefore call for different acquisition strategies, and this work enables them to coexist within one programmable system. The shared digital acquisition funnel enables this flexibility with limited additional hardware overhead.

Future work will focus on silicon characterization and experimental validation. Beyond manual configuration, the programmable CPU could enable dynamic adaptation of acquisition settings. For instance, high-resolution recordings could be analyzed to identify the minimum resolution and bandwidth needed to preserve relevant neural information, allowing subsequent acquisition with reduced requirements. Such feedback-driven adaptation would extend the architecture towards signal-adaptive acquisition.

\bibliographystyle{IEEEtran}
\bibliography{IEEEabrv,main}

\end{document}